%% file: admx_results_2009.tex
\documentclass[twocolumn,showpacs,preprintnumbers,amsmath,amssymb]{revtex4}

\usepackage{graphicx}
\usepackage{dcolumn}
\usepackage{bm}
\usepackage{subfigure}

\newcommand{\muev} {$\mathrm{\mu eV}$}

\begin{document}

\input{expected_variables.tex}

\preprint{APS/123-QED}

\title{A SQUID-based microwave cavity search for dark-matter axions}


\author{S.J. Asztalos\footnote{Currently at XIA LLC, 31057 Genstar Rd., Hayward CA, 94544.}, G. Carosi, C. Hagmann, D. Kinion and K. van Bibber}
\affiliation{Lawrence Livermore National Laboratory, Livermore, California, 94550}
\author{M. Hotz, L. Rosenberg and G. Rybka}
\affiliation{University of Washington, Seattle, Washington 98195}
\author{J. Hoskins, J. Hwang\footnote{Currently at Pusan National University, Busan, 609-735, Republic of Korea.}, P. Sikivie and D.B. Tanner}
\affiliation{University of Florida, Gainesville, Florida 32611}
\author{R. Bradley}
\affiliation{National Radio Astronomy Observatory, Charlottesville, Virginia 22903}
\author{J. Clarke}
\affiliation{University of California and Lawrence Berkeley National Laboratory, Berkeley, California 94720}

\date{\today}

\begin{abstract}

Axions in the $\mu$eV mass range are a plausible cold dark matter candidate
and may be detected by their conversion into microwave photons in a resonant cavity 
immersed in a static magnetic field. The first result from such an axion 
search using a superconducting first-stage amplifier (SQUID) is reported. The
SQUID amplifier, replacing a conventional GaAs field-effect transistor amplifier, 
successfully reached axion-photon coupling sensitivity in the band set by present axion models
and sets the stage for a definitive axion search utilizing near quantum-limited SQUID
amplifiers.
\end{abstract}

\maketitle


The axion is a hypothetical particle that may play a central role in particle 
physics, astrophysics and cosmology. Axions are pseudoscalars that 
result from the Peccei-Quinn solution to the strong CP problem \cite{PhysRevLett.38.1440,PhysRevLett.40.223,PhysRevLett.40.279}. 
Axions or axion-like particles may also be a
fundamental feature of string theories \cite{Witten}. Low mass axions ($m_a = \mu$eV-meV)
may have been produced in the early universe in quantities sufficient to account for a large
portion of the cold dark matter in galactic halos \cite{Preskill1983127,Abbott1983133,Dine1983137,ipser-sikivie}. 
These dark matter axions have extremely feeble couplings to normal matter and radiation, but 
may be converted into detectable microwave photons using the inverse Primakoff effect as 
first outlined by Sikivie \cite{PhysRevLett.51.1415,PhysRevD.32.2988}. 
Searches based on this technique are by far the most sensitive for low mass
dark-matter axions. A comprehensive dark matter axion review can be found in 
\cite{RevModPhys-Bradley}. In this Letter we describe the first results from an
axion search that uses a dc SQUID (Superconducting QUantum Interference Device), 
which offers a 2 order of magnitude
improvement in the scan rate of our search.

\begin{figure*}
\begin{center}
\includegraphics[angle=0,width=13.cm]{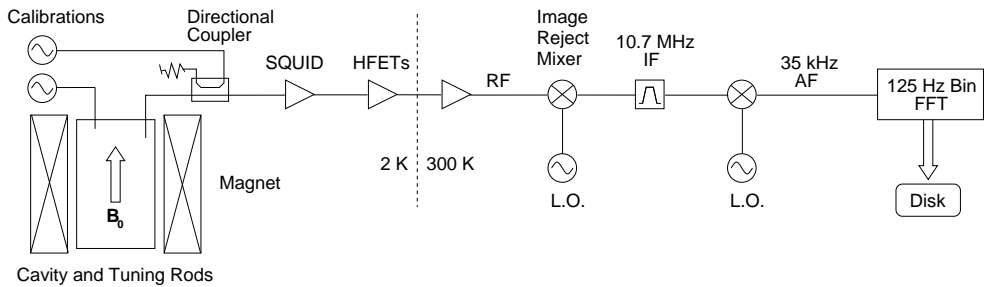}
\caption{\label{fig:ADMX_schem} Schematic of ADMX experiment. The lower left-hand sweep oscillator, which is 
weakly coupled to the cavity,
determines the resonant frequency of the TM$_{010}$ mode, while the upper left-hand oscillator allows for a reflection check in order to
critically couple the signal antenna to the cavity.} 
\end{center}
\end{figure*}

The Axion Dark Matter eXperiment (ADMX) has been running in various 
configurations at Lawrence Livermore National Laboratory (LLNL) since 1996.
The ADMX experimental configuration is sketched in Fig. \ref{fig:ADMX_schem}.
Virtual photons are provided by a 7.6 tesla magnetic field generated by
a large superconducting solenoid with a 0.5 m diameter bore. 
A cylindrical copper-plated microwave cavity is embedded in the magnet bore, and
dark matter axions passing through the cavity 
can resonantly convert into real microwave photons with energy 
$E \approx m_a c^2 + \frac{1}{2} m_a c^2 \beta^2$. With expected velocity 
dispersions of $\Delta \beta \sim 10^{-3}$ for virialized dark matter in our galaxy,
the spread in energy should be 
$\sim 10^{-6}$ or $\sim 1.2$ kHz for a 5 $\mu$eV axion. The expected power
generated by axion-photon conversions is given by \cite{PhysRevLett.51.1415,PhysRevD.32.2988},
\begin{equation}
P_a = g_{a\gamma\gamma}^2 V B_0^2 \rho_a C_{lmn} {\rm min}(Q_L,Q_a).
\label{equ:power}
\end{equation} 
Here $g_{a\gamma\gamma}$ is the coupling strength of the axion to two photons,
 $V$ is the cavity volume, $B_0$ is the magnetic field, $\rho_a$ is the
local axion dark matter density, $Q_L$ is the loaded cavity quality factor
(center frequency over bandwidth), $Q_a \sim 10^{6}$ is the axion signal 
quality factor (axion energy over energy spread) and $C_{lmn}$ is a form 
factor for the TM$_{lmn}$ cavity mode (overlap of static B field with
oscillating E field of the particular mode). In ADMX the TM$_{010}$ mode 
provides the largest form factor ($C_{010} \approx 0.69)$ \cite{Peng2000569} 
and its frequency can be moved up by translating copper-plated axial tuning
rods from the edge of the cavity to the center. Given the
experimental parameters, $P_a$ is expected to be of order $10^{-22}$ W. 
The coupling constant $g_{a\gamma\gamma} \equiv g_\gamma \alpha/\pi f_a$, 
where $\alpha$ is the fine-structure constant, $f_a$ is the ``Peccei-Quinn symmetry breaking
scale'' (an important parameter in axion theory), and $g_\gamma$
is a dimensionless model-dependent coefficient of $O(1)$. A representative choice 
within the so-called KSVZ (for Kim-Shifman-Vainshtein-Zakharov) family of 
models has $g_\gamma \sim 0.97$ \cite{KSVZ_1,KSVZ_2} while 
one particular choice within the GUT inspired DFSZ (for Dine-Fischler-Srednicki-Zhitnitshii) 
family of models has 
$g_\gamma \sim -0.36$ \cite{DFSZ_1,Dine1981199}.
Detailed experimental 
descriptions along with previous results can be found in \cite{duffy:012006} and \cite{PhysRevD.69.011101}.

The sensitivity of the detector is set by the Dicke radiometer equation 
\cite{Dicke} in which the signal-to-noise ratio is
\begin{equation}
\mbox{\it{SNR}} = \frac{P_a}{P_N}\sqrt{B t} = \frac{P_a}{k_B T_S}\sqrt{\frac{t}{B}}.
\label{equ:Dicke}
\end{equation}
Here $P_N$ is the system noise power, $k_B$ is Boltzmann's constant, $B$ is the bandwidth 
and $t$ is the integration time. The system noise temperature $T_S$
is the sum of the physical cavity temperature $T_C$ and 
the amplifier noise temperature $T_A$. 
In searching the mass range for an axion with a given coupling $g_{a\gamma\gamma}$ the 
scan rate is given by 
\begin{equation}
\frac{dm_a}{dt} \propto (B_0^2 V)^2 \cdot \frac{1}{T^2_S}
\label{equ:Scan_Speed}
\end{equation}
while, given a specific logarithmic scan rate, the smallest detectable
coupling ($g_{a\gamma\gamma}^2 \propto P_a$) is given by
\begin{equation}
g_{a\gamma\gamma}^2 \propto (B_0^2 V)^{-1} T_S.
\label{equ:Coupling_Reach}
\end{equation}

\begin{figure}
\begin{center}
\includegraphics[angle=0,width=7.0cm]{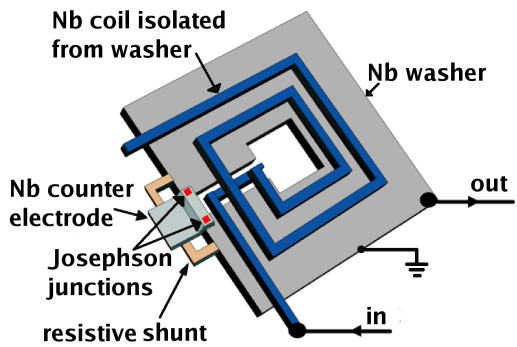}
\caption{\label{fig:squid2} Schematic of a microstrip SQUID amplifier.}
\end{center}
\end{figure}

Clearly there is a high premium on reducing $T_S$ to its lowest achievable value.
Earlier experiments used balanced GaAs heterostructure field-effect transistor (HFET)
cryogenic amplifiers built by the National Radio Astronomy Observatory (NRAO) 
\cite{RevModPhys-Bradley,Daw-Bradley-HFETs} for 
first stage amplification. 
HFET amplifiers have noise temperatures that drop as their physical temperature is lowered 
to around 10 to 20 K, at which point the noise temperature
plateaus at a value of a few K. Though extremely quiet by radio astronomy standards,
in this application their intrinsic noise of a few K severely limited the scan speed and 
resolution of the coupling constant in
previous experiments. This limitation spurred the development in the late 1990's of
replacement amplifiers for ADMX based on dc SQUIDs. 
Although dc SQUIDs have been used as amplifiers for decades 
\cite{clarke:squids}, they suffer
from severe gain roll-off at microwave frequencies due to parasitic coupling between the
input coil and SQUID washer. The SQUID amplifiers developed for ADMX are based on
a novel geometry [Fig. \ref{fig:squid2}], where the input coil is replaced by a resonant microstrip input coil \cite{Muck1998}.
The SQUID amplifier used in the axion search reported here has an in situ microwave power gain 
of $\sim$ 10 dB in the frequency range scanned.

\begin{figure}
\begin{center}
\includegraphics[angle=0,width=7.5cm]{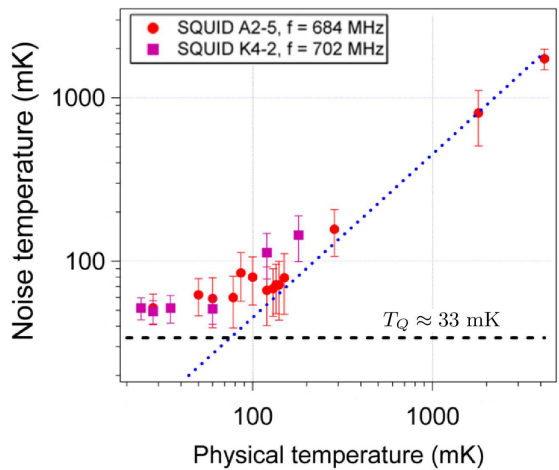}
\caption{\label{fig:squid1} Noise temperature of two representative SQUID amplifiers (with resonant frequency
\textit{f}) as a function of physical temperature. Dashed line indicates $T_Q$, the quantum noise temperature at $\approx$ 700 MHz. Dotted line has unity slope, indicating that $T_A \propto T$ in the classical regime.}
\end{center}
\end{figure}

Unlike HFET amplifiers, the SQUID amplifier noise temperature continues to drop with decreasing
temperature until it approaches the quantum noise limit ($T_Q=\hbar \omega / k_B \approx 50$ mK at 1 GHz). 
Figure \ref{fig:squid1} shows this behavior for two SQUIDs operating on resonance at 684 and 702 MHz. 
At the lowest
temperatures, their noise temperatures of $47 \pm 5$ mK are a factor of 1.4 above the quantum-limited noise
temperature of 33 mK.
Though future experiments will have dilution refrigeration to cool
the SQUID and cavity to $\sim$ 100 mK,
the current phase of the experiment used pumped liquid helium (LHe) to 
maintain cavity and SQUID temperatures of $\sim$ 2 K. 
For most of the data run, the cavity was kept under vacuum and
cooled via a small LHe reservoir fixed to the cavity top and pumped down to $\sim$ 1 torr.
The SQUID housing was thermally attached via a copper cold finger and copper strap
to this reservoir. Regions in frequency where a TE or TEM mode crossed 
the TM$_{010}$ mode were scanned by filling the cavity with superfluid LHe which shifted
the mode-crossing by $\sim$ 3\%.

Given its extraordinary flux sensitivity, placing a SQUID
amplifier in the strong fringe field of the ADMX magnet provided an additional challenge. To solve
this, the SQUID was placed $\sim$1 m above the top of the 
solenoid where the axial field has diminished to $\sim$0.5 T and inside 
a superconducting ``bucking magnet'' solenoid which canceled the fringe field
to a few 100 $\mu T$. Two nested layers of cryogenic $\mu$-metal further reduced the 
field during cool down, and the SQUID itself was placed in a superconducting, lead-plated 
housing to reject any remaining stray field. Hall sensors inside and
outside the $\mu$-metal shielding monitored the magnetic fields.

Following the SQUID were second- and third-stage 
cryogenic HFET amplifiers. These provided an additional 12 dB combined power gain, and
contributed a negligible amount to the system noise temperature. The signal was routed 
via RG-402 coaxial cable to a room-temperature post-amplifier before being 
coupled to a double-heterodyne receiver, consisting
of an image-rejection mixer with an intermediate frequency (IF) of 10.7 MHz. 
This IF stage included an eight-pole crystal filter with a 30 kHz bandwidth.
The signal was then mixed-down a second time with a doubly-balanced
mixer to an audio frequency (AF) of 35 kHz. This signal was digitized and analyzed in hardware via
fast-Fourier transform (FFT), optimized to search for the fully virialized 
axion signal. This is our medium resolution channel.
At each tuning-rod setting, 10,000 8-msec spectra at 125 Hz Nyquist resolution were
added for a total exposure of 80 s. This resulted in
a 400 point, single-sided 125 Hz resolution power spectrum. After each 80 s
acquisition, the tuning rods moved the TM$_{010}$ mode by $\sim$ 2 kHz. The
loaded $Q_L$ was remeasured before another 80 s acquisition began at the new frequency.
The overlap between adjacent spectra was such that each 125 Hz 
frequency bin had $\sim$ 25 minutes of exposure. 

A high resolution channel, not used in this analysis, is sensitive to axion spectral lines
much narrower than 125 Hz. 
In this channel, after passing through a 6.5-kHz wide passband filter, the 35-kHz signal is
mixed to an AF of 5 kHz. This is digitized and a single
power spectrum is obtained by acquiring 2$^{20}$ points over 53 s for a 
Nyquist frequency resolution of 19 mHz. Results from this channel will be described in a future 
paper. 

	Each raw power spectrum was corrected for the receiver 
input-to-output transfer function.  The frequency response of the transfer 
function is dominated by the IF crystal filter, and its effect was determined 
by an average of many spectra taken over a range of cavity frequency settings.
The remaining frequency variation of the transfer function, primarily due to frequency dependent
interactions of the cavity, transmission line and amplifier input, was removed by 
fitting and dividing each spectra by a \NPARAMETERFIT\ parameter polynomial.  
Spectra for which the chi-square of this fit (excluding peaks) was greater than \CHISQCUT\ 
were discarded as the receiver transfer function may have been poorly estimated in these cases.  

\begin{figure}
\begin{center}
\includegraphics[angle=0,width=8.0cm]{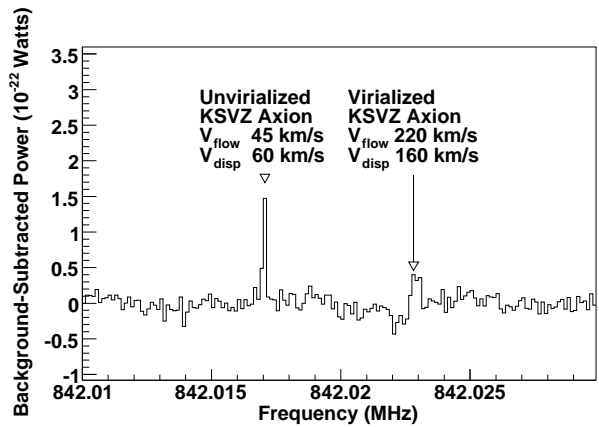}
\caption{\label{fig:example_signals} Dark matter axion signals simulated with Monte Carlo and imposed on real data for two dark matter axion distribution models (masses arbitrarily chosen).}
\end{center}
\end{figure}

\begin{figure}
\begin{center}
\thinspace
\thinspace 
\includegraphics[angle=90,width=8.0cm]{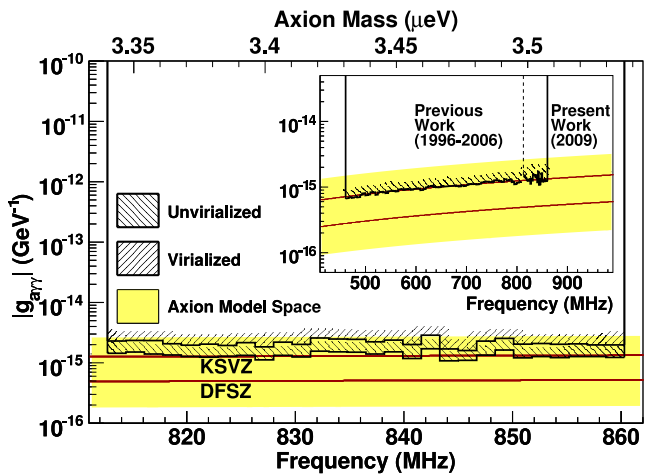}
\caption{\label{fig:limit_coupling} Axion-photon coupling excluded at the 90\% confidence level assuming a local 
dark matter density of \DMDENSITY\ for two dark matter distribution models.  The shaded region corresponds to the 
range of the axion photon coupling models discussed in \cite{PhysRevD.58.055006}.}
\end{center}
\end{figure}

Frequencies were rescanned and the power in the bins averaged until 
the expected signal-to-noise for a KSVZ axion at a 
dark matter density of \DMDENSITY\ was greater than 3.5. 
The average signal-to-noise for this run was \AVERAGESNR.
After this, bins of width 125 Hz were examined for excess 
power above the thermal power level.  Bins that contained too much measured power to 
exclude KSVZ axions were rescanned several times, and these spectra 
averaged with the previous data run. 
A characteristic of a true axion signal is that it would reappear in a rescan,
whereas statistical fluctuations or transient environmental signals would not.  
Such rescans were performed within weeks 
of the original scan, during which a putative axion signal could have shifted at most 
20 Hz due to the Earth's orbit and rotation, far smaller than the medium resolution bin width 
 \cite{PhysRevD.70.123503}.  In this run, the number of rescans agreed
with statistical expectations from thermal noise.
No signals were found to persist after the second rescan. 

	The total power and expected axion SNR were used to 
set a limit on the product of axion coupling and local dark matter density.
Two models for the axion spectral line shape were examined: completely virialized axions with a velocity dispersion of 160 km/s and a velocity relative to earth of 220 km/s, and axions with a velocity dispersion and relative velocity of 60 km/s or less, as would be predicted by a caustic model
\cite{2008PhRvD..78f3508D}
or a dark disk model
\cite{read_darkdisk_arxiv}.
Expected signals for both models superimposed on real data are shown in Fig. \ref{fig:example_signals}.  Models with lower velocity dispersions produce narrower peaks in the power spectrum, with a consequently higher SNR.  The 90\% confidence bound on axion coupling with a local dark matter density of \DMDENSITY\ is shown in Fig. \ref{fig:limit_coupling}.

	We exclude at 90\% confidence realistic axion models of dark matter, 
with a local density of \DMDENSITY\ for axion masses ranging 
3.3 \muev\ to \MAXMASSREACHED\ \muev.  This extends the excluded 
region from that covered in ref. \cite{PhysRevD.69.011101}, excluding plausible axion dark matter models from 1.9 \muev to \MAXMASSREACHED\ \muev.  
Additionally, we have demonstrated the first application of a dc SQUID amplifier 
in a high field environment 
with a noise temperature comparable to our previous runs.
In the next phase of ADMX, the SQUID and cavity will be cooled with a dilution refrigerator to 100 mK, allowing the detector
to scan over the plausible axion mass range several hundred times faster at the present sensitivity, or to be sensitive to 
even the most pessimistic axion-photon couplings over the entire axion mass range
while still scanning ten times as fast as the present detector.


This research is supported by the U.S. Department of Energy, Office of High Energy Physics under contract numbers DE-FG02-96ER40956
(Lawrence Livermore National Laboratory), DE-AC52-07NA27344 (University of Washington), and DE-FG02-97ER41029 (University of Florida). 
Additional support was provided by Lawrence Livermore National Laboratory under the LDRD program.
The National Radio Astronomy Observatory is 
a facility of the National Science Foundation operated under cooperative 
agreement by Associated Universities, Inc.
Development of the SQUID amplifier and J.C. were supported by the 
Director, Office of Science, Office of Basic Energy Sciences, Materials 
Sciences and Engineering Division, of the U.S. Department of Energy 
under Contract No. DE-AC02-05CH11231.

\bibliographystyle{h-physrev}
\bibliography{admx_results_2009}

\end{document}

%% file: expected_variables.tex
\newcommand{\MAXMASSREACHED} {3.53}
\newcommand{\AVERAGESNR} {10.4}
\newcommand{\DMDENSITY} {0.45 $\mathrm{GeV/cm^3}$}
\newcommand{\NPARAMETERFIT} {6}
\newcommand{\CHISQCUT} {1.4}

%% file: admx_results_2009.bbl
\begin{thebibliography}{10}

\bibitem{PhysRevLett.38.1440}
R.~D. Peccei and H.~R. Quinn,
\newblock Phys. Rev. Lett. {\bf 38}, 1440 (1977).

\bibitem{PhysRevLett.40.223}
S.~Weinberg,
\newblock Phys. Rev. Lett. {\bf 40}, 223 (1978).

\bibitem{PhysRevLett.40.279}
F.~Wilczek,
\newblock Phys. Rev. Lett. {\bf 40}, 279 (1978).

\bibitem{Witten}
P.~Svr\v{c}ek and E.~Witten,
\newblock J. High Energy Phys. {\bf 2006}, 051 (2006).

\bibitem{ipser-sikivie}
J.~Ipser and P.~Sikivie,
\newblock Phys. Rev. Lett. {\bf 50}, 925 (1983).

\bibitem{Preskill1983127}
J.~Preskill, M.~Wise, and F.~Wilczek,
\newblock Phys. Lett. B {\bf 120}, 127 (1983).

\bibitem{Abbott1983133}
L.~F. Abbott and P.~Sikivie,
\newblock Phys. Lett. B {\bf 120}, 133  (1983).

\bibitem{Dine1983137}
M.~Dine and W.~Fischler,
\newblock Phys. Lett. B {\bf 120}, 137  (1983).

\bibitem{PhysRevLett.51.1415}
P.~Sikivie,
\newblock Phys. Rev. Lett. {\bf 51}, 1415 (1983).

\bibitem{PhysRevD.32.2988}
P.~Sikivie,
\newblock Phys. Rev. D {\bf 32}, 2988 (1985).

\bibitem{RevModPhys-Bradley}
R.~Bradley {\em et~al.},
\newblock Rev. Mod. Phys. {\bf 75}, 777 (2003).

\bibitem{Peng2000569}
H.~Peng {\em et~al.},
\newblock Nucl. Instrum. Methods~A {\bf 444}, 569  (2000).

\bibitem{KSVZ_1}
J.~Kim,
\newblock Phys. Rev. Lett. {\bf 43}, 103 (1979).

\bibitem{KSVZ_2}
M.~Shifman, A.~Vainshtein, and V.~Zakharov,
\newblock Nucl. Phys. B {\bf 166}, 493 (1980).

\bibitem{DFSZ_1}
A.~Zhitnitskii,
\newblock Sov. J. Nucl. Phys. {\bf 31}, 260 (1980).

\bibitem{Dine1981199}
M.~Dine, W.~Fischler, and M.~Srednicki,
\newblock Phys. Lett. B {\bf 104}, 199  (1981).

\bibitem{duffy:012006}
L.~D. Duffy {\em et~al.},
\newblock Phys. Rev. D {\bf 74}, 012006 (2006).

\bibitem{PhysRevD.69.011101}
S.~J. Asztalos {\em et~al.},
\newblock Phys. Rev. D {\bf 69}, 011101 (2004).

\bibitem{Dicke}
R.~Dicke,
\newblock Rev. Sci. Instrum. {\bf 17}, 268 (1946).

\bibitem{Daw-Bradley-HFETs}
E.~Daw and R.~Bradley,
\newblock J. Appl. Phys. {\bf 82}, 1925 (1997).

\bibitem{clarke:squids}
J.~Clarke, A.~Lee, M.~M\"{u}ck, and P.~Richards,
\newblock Squid voltmeters and amplifiers,
\newblock in {\em The SQUID Handbook Vol. II: Applications of SQUIDs and SQUID
  systems}, pp. 1--93, 2006.

\bibitem{Muck1998}
M.~M\"{u}ck, M.-O. Andr\'{e}, J.~Clarke, J.~Gail, and C.~Heiden,
\newblock Applied Physics Letters {\bf 72}, 2885  (1998).

\bibitem{PhysRevD.58.055006}
J.~E. Kim,
\newblock Phys. Rev. D {\bf 58}, 055006 (1998).

\bibitem{PhysRevD.70.123503}
F.-S. Ling, P.~Sikivie, and S.~Wick,
\newblock Phys. Rev. D {\bf 70}, 123503 (2004).

\bibitem{2008PhRvD..78f3508D}
L.~D. {Duffy} and P.~{Sikivie},
\newblock Phys. Rev. D {\bf 78}, 063508 (2008).

\bibitem{read_darkdisk_arxiv}
J.~I. Read, G.~Lake, O.~Agertz, and V.~P. Debattista,
\newblock (2008), arXiv:0803.2714.

\end{thebibliography}
